\documentstyle{mn}
\begin{document}

\input{psfig.sty}
\def\lsun{{L_{\odot}}}
\def\msun{{\rm M_{\odot}}}
\def\msunyr{{\rm M_{\odot} {\rm yr}^{-1}}}
\def\msol{{\rm M_{\odot}}}
\def\te{T_{\rm eff}}
\def\rsun{{R_{\odot}}}
\def\be{\begin{equation}}
\def\ee{\end{equation}}
\def\m2i{M_{2,\rm i}}
\def\rmax{R_{2,\rm max}}
\def\ie{{\it i.e.\ }}
\def\etal{{\it et al.\ }}
\def\go{
\mathrel{\raise.3ex\hbox{$>$}\mkern-14mu\lower0.6ex\hbox{$\sim$}}
}
\def\lo{
\mathrel{\raise.3ex\hbox{$<$}\mkern-14mu\lower0.6ex\hbox{$\sim$}}
}
\def\simeq{
\mathrel{\raise.3ex\hbox{$\sim$}\mkern-14mu\lower0.4ex\hbox{$-$}}
}
\newcommand{\simgr}{\ga}
\newcommand{\simle}{\la}
\newcommand{\lta}{\la}
\newcommand{\gta}{\ga}
\input{epsf.sty}
\def\plotone#1{\centering \leavevmode
\epsfxsize=\columnwidth \epsfbox{#1}}
\def\plottwo#1#2{\centering \leavevmode
\epsfxsize=.45\columnwidth \epsfbox{#1} \hfil
\epsfxsize=.45\columnwidth \epsfbox{#2}}
\def\plotfiddle#1#2#3#4#5#6#7{\centering \leavevmode
\vbox to#2{\rule{0pt}{#2}}
\includegraphics{#1}}


\title{The violent past of Cygnus X--2}
\author[U. Kolb et al.]{Ulrich Kolb$^1$, Melvyn B.\ Davies$^2$,
Andrew King$^2$, and Hans Ritter$^3$\\ 
$^1$ Department of Physics \& Astronomy, The Open
University, Walton Hall, Milton Keynes, MK7~6AA \\ 
$^2$ Department of Physics \& Astronomy, University of Leicester,
Leicester, LE1~7RH \\
$^3$ Max--Planck--Institut f\"ur Astrophysik,
Karl--Schwarzschild-Str.~1, D--85740 Garching, Germany}

\date{MNRAS, accepted March 2000}

\maketitle

\begin{abstract}
Cygnus X--2 appears to be the descendant of an intermediate--mass 
X--ray binary (IMXB). Using Mazzitelli's (1989) stellar code we
compute detailed evolutionary sequences for 
the system and find that its prehistory is sensitive to stellar input
parameters, in particular the amount of core overshooting during the
main--sequence phase. With standard assumptions 
for convective overshooting
a case B mass transfer
starting with a $3.5\msun$ donor star is the most likely evolutionary
solution for Cygnus X--2. This makes the currently observed state rather
short--lived, of order 3 Myr, and requires a formation rate
$> 10^{-7}-10^{-6}$~yr$^{-1}$ 
of such systems in the Galaxy. 
Our calculations show that neutron star IMXBs with initially more
massive donors ($\ga 4 \msun$) encounter a delayed dynamical
instability; they are unlikely to survive this rapid mass
transfer phase.  
We determine limits for the age and initial parameters of Cygnus X--2
and calculate possible dynamical orbits of the system in a realistic 
Galactic potential, given its observed radial velocity. We find 
trajectories which are consistent with a progenitor binary on a
circular orbit in the Galactic plane inside the solar circle that
received a kick velocity $\leq 200$~km s$^{-1}$ at the birth of the neutron
star. The simulations suggests that about $7\%$ of IMXBs receiving an 
arbitrary kick velocity from a standard kick velocity spectrum 
would end up in an orbit similar to Cygnus X--2, while about $10\%$ of
them reach yet larger Galactocentric distances.
\end{abstract}


\section{Introduction}
\label{sec:intro}

The $9.84$~d period neutron--star binary Cygnus X--2
has long been regarded as an archetypal long--period low--mass
X--ray binary (Cowley et al.\ 1979, Webbink et al.\ 1983) where
nuclear expansion of a Hayashi line giant drives mass 
transfer. 
Yet recent optical photometry and high--resolution spectroscopy, while
confirming that the donor has a low mass (Casares et al.\ 1998; Orosz
\& Kuulkers 1999), unambiguously showed that the spectral type of the
optical counterpart is A$5\pm2$ (Casares et al.\ 1998), significantly
too hot for a Hayashi line donor. 

King \& Ritter (1999; hereafter KR) and Podsiadlowski \& Rappaport
(2000; PR) argued that the system must be the descendent of an
intermediate--mass X--ray binary (IMXB), with $3 - 4 \msun$ as the
likely initial donor mass. Such systems undergo a rapid mass transfer
phase, previously regarded as fatal, with transfer
rates exceeding the Eddington value by several orders of magnitude. 
For a narrow range of initial separations the systems never reach the
Hayashi line, or evolve away from it, during the subsequent phase with
slower mass transfer.

KR suggested that Cygnus X--2 is the product of a ``case B''
mass transfer sequence, where the donor star was already expanding
towards the giant branch when mass transfer  
began (Kippenhahn \& Weigert [1967]). By contrast, PR preferred an
evolution where mass transfer started while the donor was still on the
main sequence, with core hydrogen burning terminating during the mass
transfer phase. 
KR's considerations are semi--analytical and entirely
based on generalised main--sequences (Giannone et al.\ 1968), while PR
performed detailed 
binary sequences with a full stellar code. 
In Sect.~2 and 3 we use our evolutionary code to reexamine
critically the case B solution rejected by PR, 
and resolve the discrepancy between KR and PR. To see how common
a Cygnus X--2--like evolution is we consider IMXBs with still higher
initial donor masses in Sect.~4.

A further peculiarity of Cygnus X--2, addressed in Sect.~5, is its
dynamical state. The system has a measured line-of-sight velocity of
-208.6 km/s (Casares et al.\ 1998) and a Galactic
latitude of -11.32$^\circ$ and longitude of 87.33$^\circ$.
At a distance of 11.6 kpc from the sun (Smale 1998), this places
Cygnus X--2 at a Galactocentric distance of 14.2 kpc, and a distance
from the Galactic plane of 2.28 kpc.
Integrating the equations of motion in the Galactic potential  
we use Monte Carlo techniques to investigate possible trajectories of
Cygnus X--2.

\section{Constraining the prehistory of Cygnus X--2}
\label{constraints}

The observed location of the donor star of Cygnus X--2 in the 
middle of the Hertzsprung gap in the HR diagram, and its small 
mass ($0.4-0.7\msun$), suggest that the mass of the hydrogen--rich
envelope remaining above the donor's helium core is very small
($M_{\rm H}/M_{\rm He} \la 0.05$). Stars with an initial mass
$3-4.5\msun$ leave the core hydrogen burning phase with a helium core
mass $M_{\rm He}$ in this range (e.g.\ Bressan et al.\ 1993). Mass
transfer from such an 
intermediate--mass star on to a less massive neutron star is thermally
unstable and involves an initial phase of rapid mass transfer with
a highly super--Eddington transfer rate. This rapid phase lasts
roughly until the mass ratio is reversed and the donor's Roche lobe
expands upon further mass transfer. The subsequent slower transfer
phase proceeds on the donor's thermal timescale, which is essentially
given by the Kelvin--Helmholtz time of the donor when it left the main
sequence. 

PR identified two possible routes from the post--supernova binary,
i.e.\ after the formation of the neutron star, to the present system
configuration. One route is via a genuine case B evolution, the other
via an evolution they labelled `case AB'.
In the former the donor star has already left the main
sequence when it fills its Roche lobe for the first time. The mass
transfer phase is short--lived, with most of the mass transferred
at a highly super--Eddington rate. Hence the neutron star mass increase is
negligible, even if it accretes at the Eddington rate during the whole
evolution. In the case AB solution discussed by PR mass transfer
already starts during the donor's main--sequence phase. 
The system briefly detaches when core hydrogen burning
terminates, but mass transfer resumes when shell--burning is
well established. 
As the rapid mass transfer phase terminates before the
donor leaves the main sequence the subsequent slow phase proceeds on a
much longer timescale than in the genuine case B solution --- the
thermal time of the now less massive donor at the terminal
main--sequence. The transfer rate is sub--Eddington for some time and
the neutron star grows in mass. 

Given this we explore three different prescriptions to project the
evolution of Cygnus X--2 backwards in time:  

(Model 1) To represent a case B evolution we assume that $M_1=$const., and
that any material lost from the system carries the specific 
orbital angular momentum of the neutron star. Then the
initial orbital  separation $a_i$ is
\be
     a_i  =  a \, \frac{M_1+M_2}{M_1+M_{2i}}
               \left(\frac{M_2}{M_{2i}}\right)^2 \exp\left
               ( \frac{2(M_{2i}-M_2)}{M_1}\right) ,
\ee
where $M_{2i}$, $M_2$ denote the initial and present donor mass, and $a$
the present separation (e.g.\ KR). This solution is valid only if
$a_i > a_{\rm 
TMS}$, the orbital separation of a binary with a Roche--lobe filling
terminal main--sequence (TMS) star of mass $M_{2,i}$. 

(Model 2) The same applies to a case AB evolution, except that
we require $a_i < a_{\rm TMS}$.

(Model 3) To account for a possible prolonged phase
with sub--Eddington mass transfer in a case AB sequence we allow  
mass transfer to be conservative (total binary mass and orbital angular
momentum is constant) 
for the last part of the evolution. We assume that the neutron star
mass at birth was $1.4 \, \msun$, and that the present neutron star
mass is $> 1.4 \, \msun$. Hence the evolution {\em backwards} in time
consists of two branches. During the conservative phase, characterised
by $a M_1^2 M_2^2=$constant, $M_1$ {\em reduces} from its present value to
$1.4 \, \msun$. The second phase is calculated with constant neutron
star mass ($1.4 \msun$), as in models 1 and 2 above.

\begin{figure}
\plotone{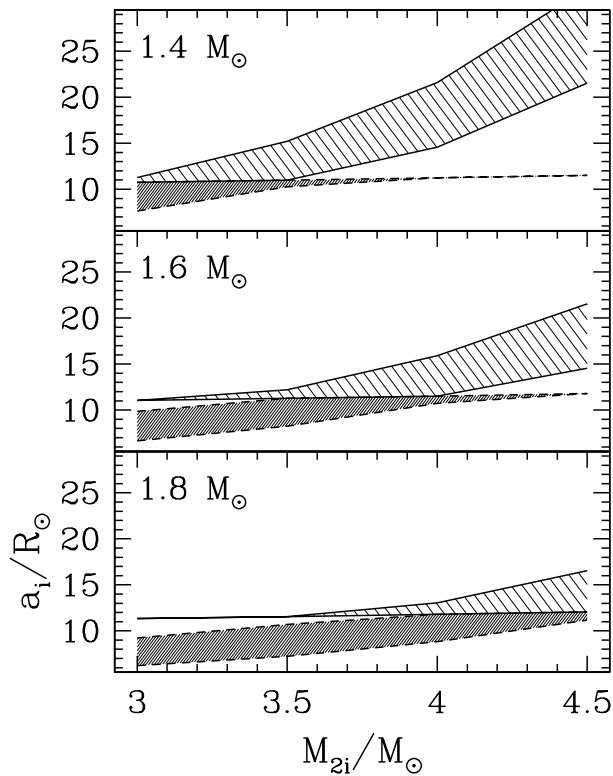}
\caption{
Allowed range for the initial orbital separation $a_i$ as a function
of initial donor mass $M_{2i}$, for three different fixed
neutron star masses, as labelled. 
The hatched areas correspond to case B evolution (model 1; wide spacing)
and case AB evolution (model 2; narrow spacing), assuming constant neutron
star mass. The parameter space for case AB solutions with model 3
assumptions is so small that it is not shown. The almost horizontal
boundary at $a_i = a_{\rm TMS} \simeq 11 \, \rsun$ separating case B
and case AB solutions represents the terminal main sequence. $a_{\rm
TMS}$ was calculated using main--sequence models by 
Dominguez et al.\ (1999). See text for details.
\label{fig0}
}
\end{figure}

The present system parameters of Cygnus X--2 are: orbital period
$P_0=9.844$~d, mass ratio $q=0.34\pm0.04$ (Casares et al.\ 1998), and
neutron star mass $M_1 = 1.4-1.8\msun$. This mass range accommodates
the canonical neutron star mass at birth, as well as the 
estimate by Orosz \& Kuulkers (1999) based on modelling
the observed ellipsoidal variations.
If we adopt a specific value $M_1$ for the present neutron star mass,
then the present donor mass is constrained to the range $0.30 M_1 \la
M_2 \la 0.38 M_1$ by the observed mass ratio.
This translates into a range of allowed initial separations, as shown 
by the hatched regions in Fig.~\ref{fig0}, for case B (model 1, wide
spacing) and case AB (model 2, narrow spacing) evolution. 
The parameter space for case AB solutions with model 3
assumptions is tiny and hence not shown in the figure.
For conservative mass transfer (model 3) the orbital separation of
Cygnus X--2 {\em today} increases less steeply with decreasing donor mass
than for the isotropic wind case (model 2). Hence the separation for
model 3 sequences describing the {\em past} 
evolution of Cygnus X--2 is generally larger than for
the corresponding model 2 sequences. In 
particular, model 3 solutions require larger initial separations. 
Most of them are in conflict with the limit $a_i<a_{\rm TMS}$, i.e.\
inconsistent with the assumption that the donor was on the main
sequence when mass transfer started.

More generally, if the mass lost from the system in sequences with
$M_1=$~const.\ carries more (less) specific angular momentum than that
of the neutron star, the required initial separation is larger
(smaller) than estimated by (1). A somewhat higher loss seems more
likely than a smaller loss, hence this reduces the
parameter space available for case AB solutions. 

Even if Fig.~\ref{fig0} indicates a viable solution it is not clear if
the corresponding evolutionary sequence reproduces Cygnus
X--2. Although the donor will have the observed mass (and hence radius)
at the observed orbital period $P_0$, the effective temperature is of
course not constrained by the above considerations. To check this
we need a detailed binary sequence with full stellar models, as
presented in the next section. In particular, it appears that the
case AB sequences described by PR require the donor star to be already
fairly close to the end of core hydrogen burning at turn--on of
mass transfer.
This is likely to narrow the parameter space for case AB sequences
even further.   

We note that the limits shown in Fig.~\ref{fig0} depend via $a_{\rm
TMS}$ 
somewhat on the stellar input physics, in particular on the amount of
convective core overshooting (cf.\ the discussion in the next
section).

\section{Model calculations}
\label{sequences}

We calculated several detailed early massive case B binary evolution
sequences, using Mazzitelli's stellar code in its 1989 version (see
Mazzitelli 1989, 
and references therein, for a summary of the input physics). Mass
transfer was treated as in Kolb (1998). Table~\ref{tab0} summarises
initial and final system parameters. The neutron star mass was $1.4
\msun=$~const.\ in each case, with angular momentum loss treated as in
model~1 above. Specifically, we present in detail the evolution of a
system with initial donor mass $M_2=3.5 \msun$ and orbital separation
$a_i=13.3 \rsun$ (Sequence I) and $11.1 \rsun$ (sequence II).  

\begin{figure}
\plotone{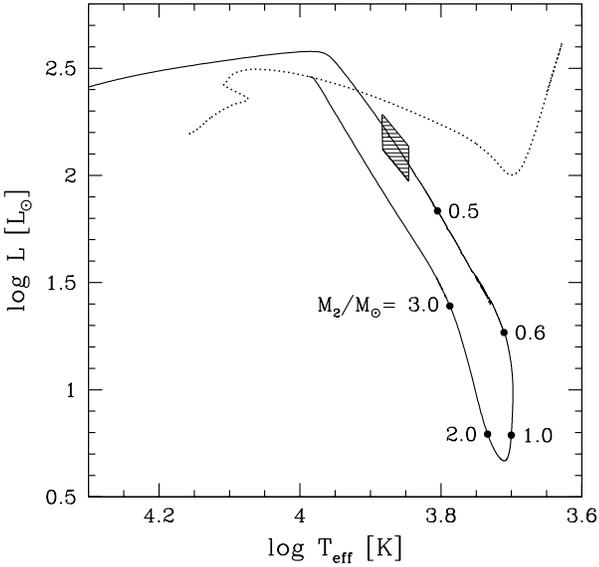}
\caption{
HR diagram showing sequence I (solid) and a track of a $3.5 \msun$
single star (dotted), calculated without core overshooting. Labels
along the track indicate the donor mass. The shaded
error box marks the observed location of Cygnus X--2 (lower and upper bound
correspond to donor masses $0.4\msun$ and  $0.7\msun$, respectively).
\label{fig1}
}
\end{figure}

\begin{figure}
\plotone{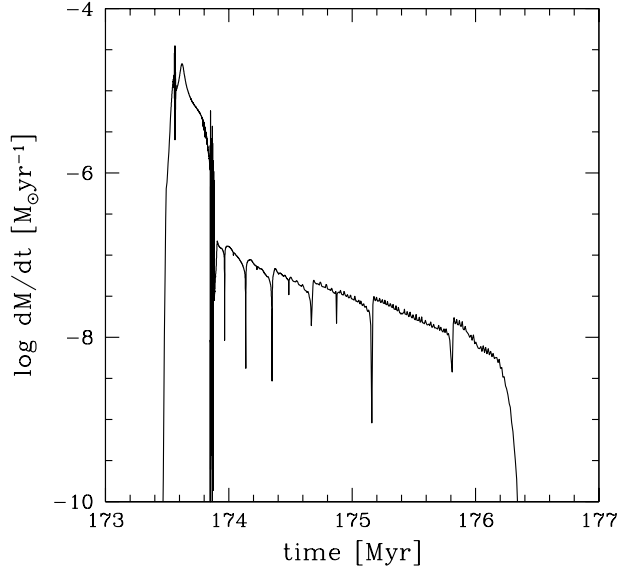}
\caption{
Mass transfer rate $\dot M$ vs time (elapsed
since the beginning of core hydrogen burning) for sequence I. 
Wiggles reflect numerical noise due to the explicit treatment of mass
transfer. The instabilities are particularly strong at the end of the
rapid phase close to $t=173.8$~Myr. 
The semi--regular dips of $\dot M$ during the slow phase are due to
sudden changes of the stellar radius, caused by a discontinuous
change of the inner boundary of the convective envelope.
\label{fig2}
}
\end{figure}

\begin{figure}
\plotone{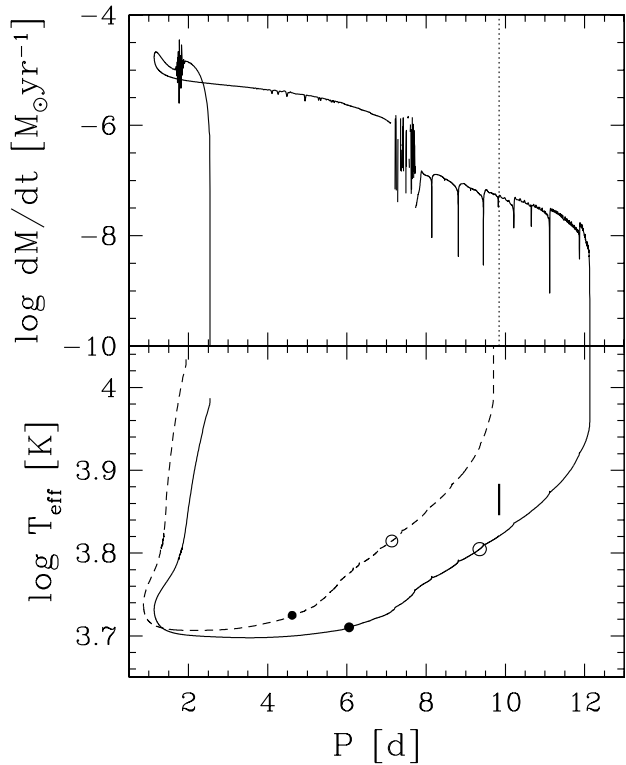}
\caption{
{\em Upper panel:}
Mass transfer rate $\dot M$ vs orbital period $P$ for sequence I. The 
numerical noise at the end of the rapid phase ($7 < P/{\rm d} < 8$)
is suppressed. The dotted vertical line marks the period
of Cygnus X--2. 
{\em Lower panel:}
Effective temperature vs $P$, for sequence I (solid)
and II (dashed). The filled and open circle on the tracks mark where
the donor mass is $0.6 \msun$ and $0.5 \msun$, respectively. The
short bar indicates the location of Cygnus X--2. 
\label{fig3}
}
\end{figure}

The figures \ref{fig1}-\ref{fig3} confirm the general behaviour 
of case B evolutionary sequences as described above. Sequence I
gets very close to the observed state of Cygnus X--2, although
it fails to fit all observed parameters simultaneously with high
accuracy. The
evolutionary track in the HR diagram passes through the Cygnus X--2 error
box, but 
at a period slightly longer than the observed value $P_0=9.844$ d. This 
can be seen in Fig.~\ref{fig3}, which shows that at $P_0$ the sequence
I donor is somewhat too cool, while in sequence II it is 
somewhat too hot. Hence the initial separation for a simultaneous fit
of $P$, $T_{\rm eff}$ (and $L$) lies between $11.1 \rsun$ and
$13.3 \rsun$. The transfer rate at $P_0$ for sequence I is $5
\times 10^{-8} \msunyr$, a factor $2-3$ higher than the Eddington
rate. 

Observational estimates place the actual accretion rate in Cygnus
X--2 close to the Eddington limit. With no evidence of 
significant mass loss from the system we expect that
the transfer rate is also of the order of the Eddington rate.
As the transfer rate in the model
sequence is set by the thermal time of the progenitor star, a somewhat
smaller initial secondary mass (e.g.\ $3.0\msun$) would give a lower
rate; this has already been noted by PR.

We note the following main differences between our sequence I and the
case B sequence calculated by PR: \\
(1) PR chose $21.3 \rsun$ as the initial separation. This leads
    to the combination $1.4+0.66 \msun$ for the component masses at
    $P=P_0$, i.e.\ to a mass ratio $0.47$, significantly larger than
    observed. Hence the PR case B sequence represents a poor fit for
    Cygnus X--2. \\ 
(2) More importantly, the mass transfer rate in the slow phase of
    sequence I decreases from $10^{-7} \msunyr$ to $10^{-8} \msunyr$
    over a period of 2.5 Myr, while in PR's model the transfer rate is
    always larger than $10^{-7} \msunyr$ (which is inconsistent with
    observations). 
    A similar calculation by Tauris et al.\ (2000) with
    essentially the same stellar code as the one used by PR gives
    a similarly high rate in the slow phase. 

The initial separation for our sequence II is essentially
the same as the one used by PR for their case AB evolution
($11.5\rsun$, corresponding to a stellar radius $5.6\rsun$ of the
donor at turn--on
of mass transfer). 
The most important difference between the two calculations is
the degree of convective overshooting assumed during the
main--sequence phase: none in our models, a very strong one in PR's
models. Hence PR's main--sequence band is much wider, with $6.6\rsun$
as the maximum radius $\rmax$ that a $3.5 \msun$ star    
reaches during core hydrogen burning, compared to $3.6 \rsun$ in our
case. A moderate extent of core overshooting is favoured in the
literature, giving $\rmax = 5.0 \rsun$ (Schaller et al.\ 1992), $5.7
\rsun$ (Bressan et al.\ 1993) and $4.8 \rsun$ (Dominguez et al.\ 1999)
for a $3.5 \msun$ star with solar composition. 
Schr\"oder et al.\ (1997) --- who use the same code and input physics
as PR --- prefer a rather efficient overshooting leading to  values up
to $\rmax = 7.1 \rsun$, while in a 
subsequent paper (Pols et al.\ 1997) the same authors concluded that
$\rmax = 5.9 \rsun$ (which corresponds to their overshooting parameter 
$\delta=0.12$) provides the best overall fit to observations.  
This is only barely larger than the donor's radius at turn--on of
mass transfer in PR's case AB solution.  
Hence standard assumptions about overshooting favour a case B solution
for Cygnus X--2 over a case AB solution.  

We conclude that a case B solution is a viable fit for the
evolutionary history of Cygnus X--2. Given the parameter space
limitations for a case AB evolution it does appear as the more likely
solution for Cygnus X--2, even though this implies that the presently
observed state of Cygnus X--2 is rather short--lived, of order several
million years.

PR pointed out that the surface composition of the donor in Cygnus
X--2 would be significantly hydrogen--depleted ($X \simeq 0.1$) and
show signs of CNO--processing if it had undergone a case AB evolution.
Unfortunately, this does not distinguish unambiguously between a case AB
and case B evolution, as the same is in principle true for a donor
that had undergone a case B mass transfer. The models in our sequence I
close to the position of Cygnus X--2 have a surface hydogen mass
fraction of $X=0.29$, while C/N and O/N are close to the 
equilibrium values for CNO burning at $2\times 10^7$~K.

\begin{table}
\begin{tabular}{llllll} \hline\hline
\noalign{\medskip}
$M_{2i}/\msun$ & $P_i/$d & $a_i/\rsun$ & $M_{2f}/\msun$ & $P_f/$d &
comment \\ 
\noalign{\medskip}
\hline
\noalign{\medskip}
3.5 & 1.94 & 11.1 & 0.44 & 9.70  & Sequence II \\
3.5 & 2.54 & 13.3 & 0.45 & 12.13 & Sequence I \\
3.5 & 3.13 & 15.3 & 0.45 & 14.65 & \\
\noalign{\medskip}
3.75 & 4.23 & 19.0 & 0.50 & 13.24 & \\ 
\noalign{\medskip}
4.0  & 1.25 & 8.55 & 3.99 & 1.24  & runaway \\
4.0  & 6.43 & 25.5 & 3.00  & 0.443 & runaway \\
\noalign{\medskip}
5.0  & 4.86 & 22.4 & 4.34  & 2.19  & runaway \\
\noalign{\medskip}
\hline
\end{tabular}
\caption{\label{tab0} Donor mass $M_{2f}$ and orbital period $P_f$
when mass transfer ceases for sequences with different initial donor
mass $M_{2i}$ and initial orbital period $P_i$ (initial orbital
separation $a_i$). Sequences 
labelled ``runaway'' encounter the delayed dynamical instability;
$M_{2f}$ and $P_f$ refer to the system at this point.
}
\end{table}

\section{Sequences with initially more massive donor stars}

The calculations discussed so far represent early case B mass transfer
solutions for systems with a moderate initial mass ratio,
$q=M_2/M_1\ga2.5$. Systems with smaller mass ratio ($q\la1$) have no
initial rapid mass transfer phase and can be understood
semi--analytically (see e.g.\ Kolb 1998, Ritter 1999), while the fate
of systems with yet larger initial mass ratio $\ga 3$ is
unclear. Hjellming (1989) pointed out that sustained 
thermal--timescale mass transfer can lead to a delayed transition to
dynamical--timescale mass transfer. The reason for this is that the
adiabatic mass--radius index of initially 
radiative stars decreases significantly when the radiative envelope is
stripped rapidly. This causes the Roche lobe to shrink faster than the 
star. Estimates for the critical initial
mass ratio which just avoids the delayed dynamical instability give 
$q\simeq 3-4$ (Hjellming 1989; Kalogera \& Webbink 1996), although
none of these are based on self--consistent mass transfer
calculations. 

This critical value is relevant for identifying possible descendants
of neutron--star systems undergoing early massive case B mass transfer.
These should appear as binary millisecond pulsars lying significantly
below the orbital period--white dwarf mass relation found by 
Rappaport et al.\ (1995; see also Tauris \& Savonije 1999 for
calculations with updated input physics) 
for systems descending from Hayashi line low--mass
X--ray binaries. Obvious candidates are systems with a  
fairly high--mass white dwarf but short orbital period. The most
discrepant system is B0655+64 ($P=1.03$~d, $M_{\rm WD} \ga 0.7$; cf.\
KR). The fairly massive white dwarf implies a donor mass $\ga 5\msun$
in the progenitor binary.

We performed test calculations similar to our sequences I and II, but with
initial donor mass $3.75 \msun$, $4.0 \msun$ and $5.0 \msun$. While
the $3.75 \msun$ sequence was stable throughout, the $4.0 \msun$ and 
$5.0 \msun$ sequences encountered runaway mass transfer (where
we stopped the calculations) rather early in the rapid mass transfer
phase (see Tab.\ref{tab0}).
An additional test sequence with a $3 \msun$ donor near the end of
core hydrogen burning and initial mass
ratio 4 (i.e.\ primary mass $0.75 \msun$, assumed constant, as above)
encountered runaway mass transfer at a donor mass $2.6 \msun$. This is
in perfect agreement with Hjellming's 
prediction, based on his Fig.~IV.1 and a Roche lobe curve
corresponding to $q=4$.    

Unless a neutron star binary can survive even such a
dynamical--timescale mass transfer, the phase space available for
a Cygnus X--2--like 
evolution is severely limited by this upper limit on the initial donor
mass. Our calculations suggest that if most neutron stars form with
$1.4 \msun$ the maximum white dwarf mass in an endproduct of early
massive case B evolution is $\la 0.55 \msun$. If this is true, neither
B0655+64 nor the recently discovered system J1453-58 ($P=12.422$~d,
$M_{\rm WD} \ga 0.88 \msun$; cf.\ Manchester et al.\ 1999) can have
formed in this way. 

The occurrence of the delayed dynamical instability is
intimately linked to the internal structure of the donor
star. Therefore it is not surprising that the maximum initial donor
mass for early case B mass transfer, just 
avoiding this instability, depends on the stellar input physics. 
Using an updated version of Eggleton's stellar code (see e.g.\ Tauris 
\& Savonije 1999), Tauris et al.\ (2000) found $5 \, \msun$ for this
limit, with a corresponding maximum white dwarf mass $0.8 \, \msun$ 
in millisecond pulsar binaries formed in this way.

\section{The trajectory of Cygnus X--2}

As described in the introduction, Cygnus X--2 is in an unusual 
dynamical state.
In this section we investigate possible trajectories of Cygnus X--2,
given the observed line--of--sight velocity and the constraints on the
evolutionary state as derived in Secs.~2 and 3.

In Fig. 5 we plot the cumulative  velocity distribution 
for systems with initial separations of $d = 15R_\odot$, i.e.\
immediately after circularisation of the binary orbit following the
SN explosion producing the neutron star. 
The secondary is taken
to have a mass $M_2=3.5$ M$_\odot$ while the mass of the helium star
prior to the supernova is $M_1=5.0$ M$_\odot$. 
We consider two different
distributions for the kick velocity imparted to the neutron star;
namely those by Hansen \& Phinney (1997) and Fryer (1999). 
Throughout the following section, when we refer to kick velocities
we mean the kick imparted on the binary allowing for the effects of
mass--loss and an asymmetric supernova, and not simply the 
kick imparted on the neutron star from the latter.

Given its present position and
velocity, we integrate the trajectory of Cygnus X--2 backwards to the
birth of the neutron star  
using a model for the Galactic potential suggested by Paczy\'nski
(1990) (reviewed in the Appendix). From sections 2 and 3 it is clear
that the age $t$ of the neutron star is essentially the main--sequence
lifetime of the progenitor star of the present donor. To account for
the allowed range of the initial donor mass and for uncertainties from
the width of the main--sequence we choose $150 < t/{\rm Myr} < 275$.
As only the line-of-sight velocity, ${\bf v}_{los}$, is known, 
we considered a set of trajectories where the 
current velocity was given by 

\begin{equation}
{\bf v} = {\bf v}_\odot + {\bf v}_{los} + \alpha {\bf v}_1 + \beta {\bf v}_2  
\end{equation}
where ${\bf v}_{los}$, ${\bf v}_1$, and ${\bf v}_2$ are mutually orthogonal,
and ${\bf v}_1$ has a zero component in the $z$ direction. 
When computing trajectories, it is most convenient to work in units
related to the mass and size of the galaxy. 
As a natural unit of velocity we use 207~km/s, the Kepler velocity at  
1~kpc distance from a point mass $10^{10}$ M$_\odot$.
In the above equation 
${\bf v}_1$, and ${\bf v}_2$ are given unit lengths in our code units.

\begin{figure}
\plotone{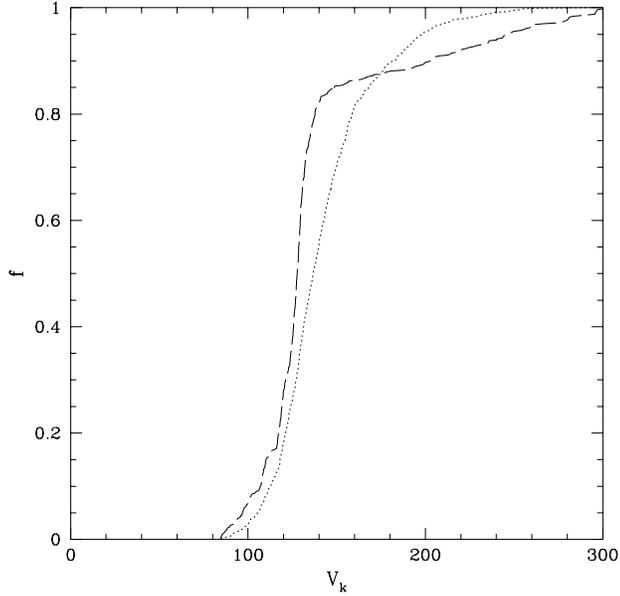}
\caption{The cumulative velocity distribution (in km/s) for systems 
with initial separations, $d = 15R_\odot$. 
This is the velocity of the centre--of--mass of the binary as opposed
to the kick of the neutron star itself. Dotted line
is computed from the neutron star kick distribution of 
Hansen \& Phinney (1997), whereas the dashed line is
for the kick distribution of Fryer (1999).}
\end{figure}

\begin{figure}
\plotone{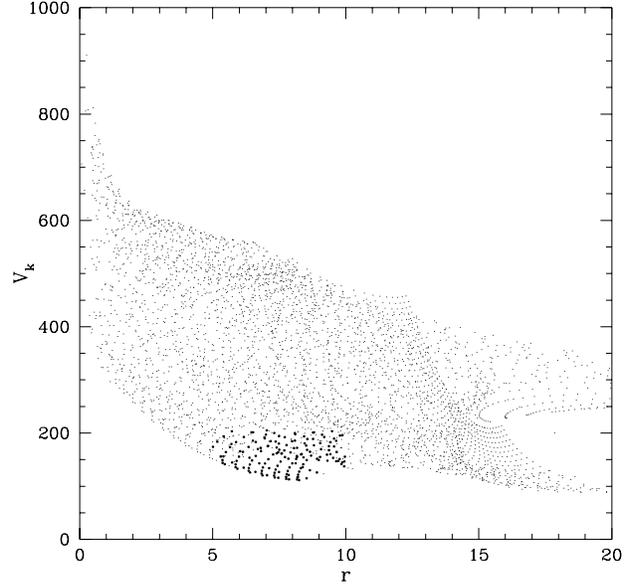}
\caption{The kick velocity, $V_{\rm k}$ (in km/s),
 of binaries as a function
of formation radius, $r$ (in kpc). In the case of the larger dots,
$150 < t < 275$, and $5 < r < 10$.} 
\end{figure}

\begin{figure}
\plotone{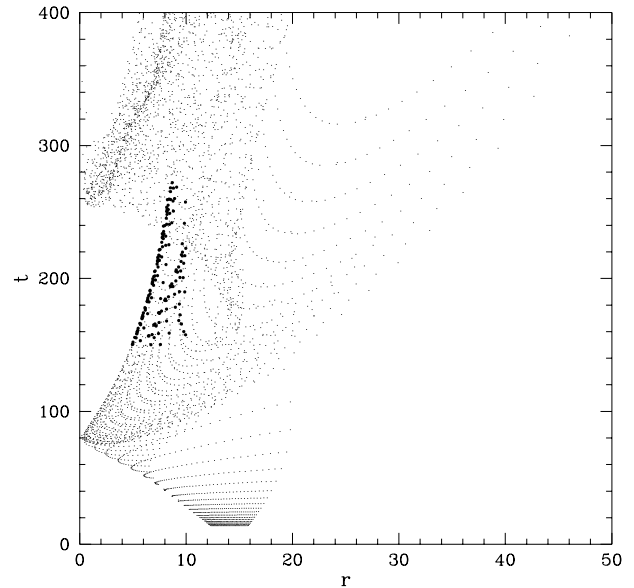}
\caption{The age of formation, $t$ (in Myr), of binaries as a function
of formation radius, $r$ (in kpc). For the larger dots,
$150 < t < 275$, and $V_{\rm k} \le 200$ km/s.}
\end{figure}

We integrated the equations (A9) for a range of values of $\alpha$
and $\beta$ and noted the radius, $r$, and time, $t$,
whenever the  trajectory
cut the Galactic plane. We also computed the kick velocity,
$V_{\rm k}$  the system
must have had if the neutron star has formed
at that time, assuming the system had previously 
been in a circular orbit in the Galactic plane.

In Fig.~6 we plot the values of $r$ and $V_{\rm k}$
for all values of $\alpha$ and $\beta$. 
We note that for $r \lo 10$ kpc,
140 km/s $\lo V_{\rm k} \lo$ 600 km/s. This range of required values of
$V_{\rm k}$ should be compared with that expected given the neutron star 
kick distributions of Hansen \& Phinney (1997) and Fryer (1999)
in Fig. 5. It is clear from Fig.~5 that the expected value of $V_{\rm k}$
is far lower than that required for many of the trajectories shown in Fig.~6.
In other words, in a large fraction of systems formed at a radius $r \lo
10$ kpc, the trajectory of the binary will confine the system to the inner
regions of the Galaxy; Cygnus X--2 must have been relatively unusual in
reaching out beyond the solar circle. 

We plot the time of formation, $t$, as a function
of $r$ for all values of $\alpha$ and $\beta$ in Fig. 7. As in Fig. 6, 
the data points where $150$ Myr $< t< 275$ Myr and $V_k \le 200$ km/s
are plotted as larger dots. From the work of Sections 2 and 3 these
are the values most likely to be  
applicable to a progenitor of Cygnus X--2. We note from Fig. 7 that the binary
is unlikely to have originated within 5 kpc of the Galactic centre as
trajectories cutting the Galactic plane at such small radii do so 
at the wrong time. Fig. 6 also shows they
would require unreasonably large kicks.

In Fig. 8 we plot the values of $\alpha$ and $\beta$ for those
trajectories which cut the Galactic plane $t$ Myr ago, where
$150 < t < 275$, with system kick velocities required to
take the binary from a circular orbit restricted
by  $V_k \le 200$ km/s. Values of $\alpha$ and $\beta$
besides those shown in Fig. 8 were considered, but none satisfied
the above conditions. We see that
only a small number of possible trajectories are credible with these
restrictions applied.
Here we note the effect of the Galactic rotation. For $\alpha
\lo 0.4$ the trajectories are in the opposite direction to the Galactic
rotation. In other words the kick the system receives on formation
will oppose its initial velocity in a circular
orbit, and will therefore have to be larger than for
a similar trajectory in the same direction as the Galactic
rotation. Here the effect of the kick is boosted  by the initial velocity 
of the circular orbit.

\begin{figure}
\plotone{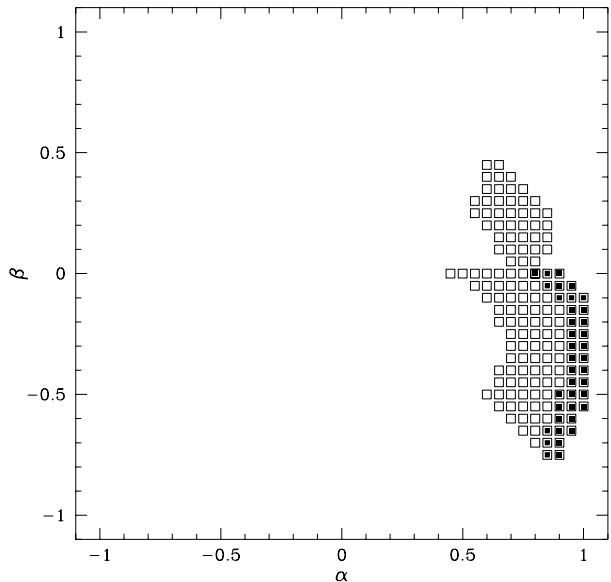}
\caption{The values of $\alpha$ and $\beta$ producing trajectories
where the binary was ejected from the Galactic plane from within a
radius, $r \le 10$ kpc. The system is assumed
to have received a kick, $V_{\rm k} \le
200$ km/s, and the ejection to have occurred a time $t$ Myr ago, where
$ 150 < t < 225$ (open squares) and $225 < t < 275$ (filled squares).}
\end{figure}

\begin{figure}
\plotone{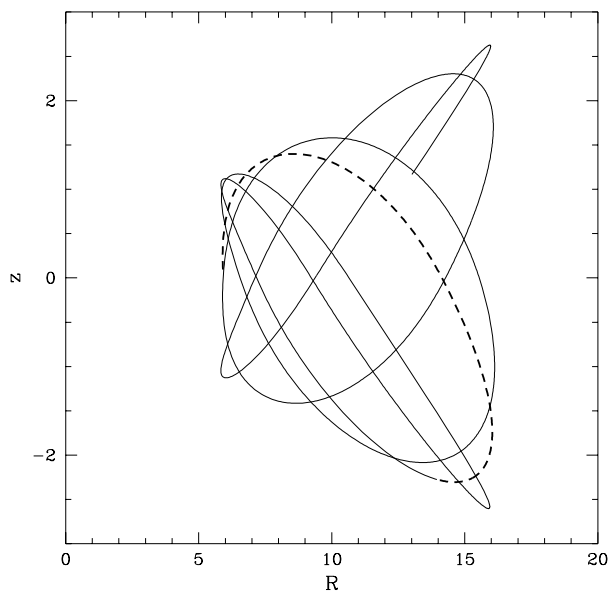}
\caption{Side view of the Galaxy (its plane is the line $z=0$), showing
the trajectory of the binary for $\alpha=0.75$ and $\beta=-0.2$.
The dashed curve shows the path from formation to the present day,
and the solid curve shows the path that the system will follow over the
next one billion years.}
\end{figure}

\begin{figure}
\plotone{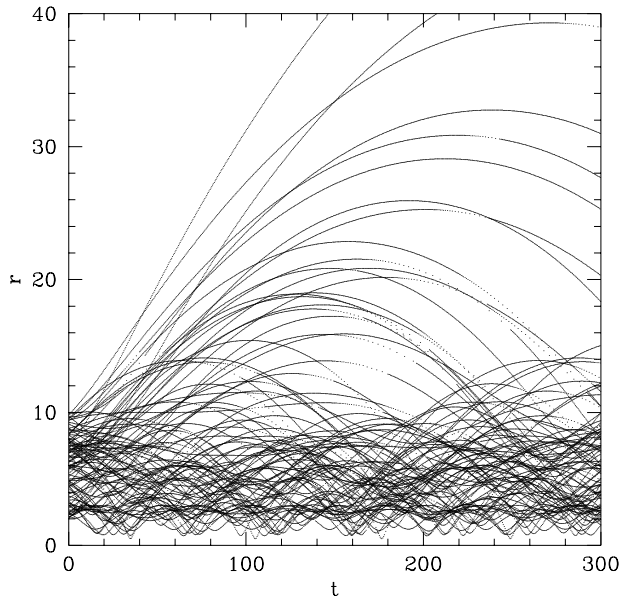}
\caption{The Galactocentric distance (in kpc) reached
by the binary as a function of time (in Myr) for 100 randomly-picked
initial conditions.}
\end{figure}

For illustration, 
in Fig. 9 we plot the trajectory ($R$ and $z$) for $\alpha = 0.75$ and
$\beta = -0.2$. In this case, the trajectory cuts the Galactic plane on
two occasions within the last 250 Myr, the first occasion being 
95 Myr ago, at a radius, $R_{\rm cut} = 14.1$ kpc, and the second
occasion being 172 Myr ago, at a radius , $R_{\rm cut} = 5.9$ kpc. 
If the system was formed in the disc at this latter time, and
was initially travelling on a circular orbit, it would have had to
receive a kick, $V_{\rm k} \sim 130$ km/s, which sits comfortably
within the range given in Fig. 5.

Thus far we have demonstrated that there are viable trajectories
for Cygnus X--2 compatible with the observed line-of-sight 
velocity. We now ask the reverse question: what fraction of binaries
originating at some radius will produce systems resembling Cygnus X--2?
Given that we expect systems to have originated within the solar
circle,
as most massive stars are located within 10 kpc of the Galactic
centre,
we might suspect that Cygnus X--2 is relatively unusual in being 
at a larger distance from the Galactic centre and significantly away from
the Galactic plane. By Monte Carlo simulation we were able to 
produce a large number of systems at various initial radii, 
assumed to be initially on circular orbits,
with kick velocities drawn from the distributions given in Fig. 5,
and follow their trajectories in the Galactic potential. Investigation
demonstrated that in determining whether a particular trajectory would
produce a Cygnus X--2 like system, the maximum radius reached by the system,
$R_{\rm max}$ was a good diagnostic. This is illustrated in Fig. 10 
where we plot the radii of 100 binaries as a function of time where
the initial trajectories are drawn randomly, from the kick distribution 
of Hansen and Phinney
(see Fig. 5), and choosing the initial radius to be between
2 kpc and 10 kpc. 
The trajectories fall into two categories: those where the binary remains
at a radius similar to that at which it was located initially, and those
where the binary is ejected significantly into the Galactic halo, reaching
maximum radii of $\go 10$ kpc. Cygnus X--2 clearly belongs to the latter
category. In order for a system to be at a radius today similar to that
of Cygnus X--2, we require $15 \lo R_{\rm max} \lo 20$ kpc, providing the
system originated in the Galactic disc somewhere within 10 kpc of the
Galactic centre. A number of systems will be ejected to even larger
radii. In such cases a binary of age similar to Cygnus X--2 would   
still be on the outward bound portion of its orbit.

In Fig. 11 we plot $R_{\rm max}$ as a function of initial radius $R$. In 
this case the initial radius was chosen
randomly from 2 kpc to 10 kpc, and the kick velocity drawn from
the distribution given by Hansen and Phinney (see Fig. 5). From this figure
we note that Cygnus X--2--like objects may be produced when $r \go 5$ kpc,
and that the relative frequency is relatively independent of formation
radius. Systems having larger values of $R_{\rm max}$ will also be produced,
the frequency increasing with radius of formation. The relative frequency
for Cygnus X--2--like systems and those on longer period orbits is plotted
as a function of $R$ in Fig. 12.

Assuming that the formation rate of binaries scales as 
the surface density  of stars, which is given by $\Sigma \sim
\Sigma_0 e^{-r/3{\rm kpc}}$ (see Binney \& Merrifield 1999), 
we computed the relative number
of Cygnus X--2--like systems which would be produced by integrating 
over the entire disc within the solar circle. 
The fraction of systems resembling Cygnus X--2 ($f_{\rm cyg}$)
and the total fraction of systems which will travel significantly
further out than  their formation radius ($f_{\rm out}$) are listed 
in Table~\ref{tab1} as a function
of the initial separation within the post-supernova binary, $d$, the
primary helium star mass {\sl prior} to the supernova, $M_1$, and the mass
of the secondary, $M_2$.
We found that $\sim 7$\%
of all binaries will produce Cygnus X--2--like binaries on trajectories
that would place them at a radius similar to Cygnus X--2 today. A further
$\sim 5 - 15$\% will be further out than Cygnus X--2, 
whilst the remainder will
be located closer to the Galactic centre. These results apply equally
to both velocity distributions plotted in Fig.~5.

\begin{figure}
\plotone{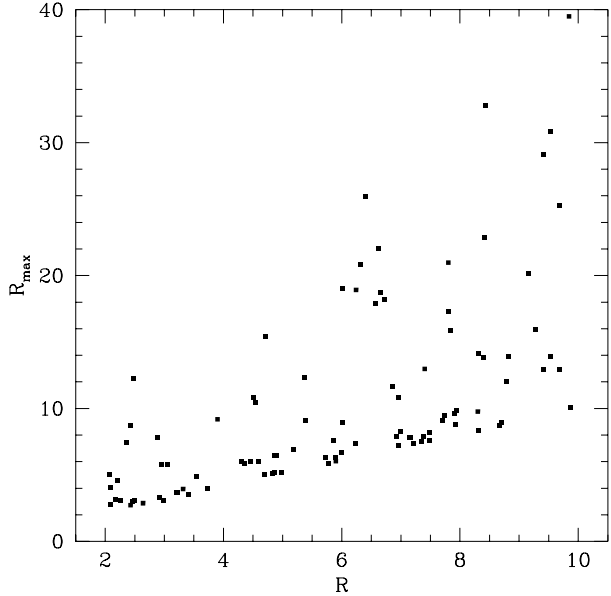}
\caption{The maximum Galactocentric distance reached
by the binary, $R_{\rm max}$, as a function of
formation radius, $R$. One hundred different trajectories are
considered here. The initial radius was chosen
randomly from two to ten, and the kick velocity drawn from
the distribution given by Hansen and Phinney (see Fig. 5).}
\end{figure}

\begin{figure}
\plotone{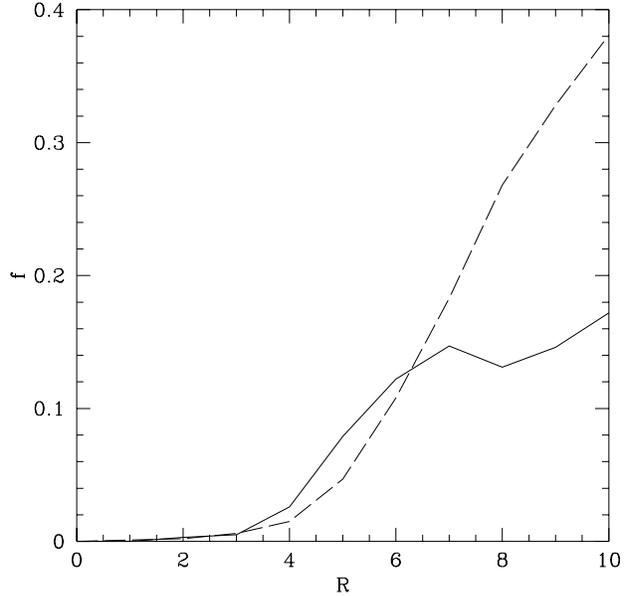}
\caption{The fraction of systems on similar trajectories to Cygnus X--2 
(solid line), and those on longer period orbits (dashed line)
as a function of formation radius, $R$ (in pc).}
\end{figure}

\begin{table}
\medskip
\begin{tabular}{lllll} \hline\hline
\noalign{\medskip}
$d/R_\odot$ & $M_1/M_\odot$ & $M_2/M_\odot$ & $f_{\rm cyg}$ & $f_{\rm out}$ \\
\noalign{\medskip}
\hline
\noalign{\medskip}
10.0 & 5.0 & 3.5 & 0.07 & 0.17 \\ 
15.0 & 5.0 & 3.5 & 0.06 & 0.15  \\ 
20.0 & 5.0 & 3.5 & 0.06 & 0.13  \\ 
25.0 & 5.0 & 3.5 & 0.06 & 0.11  \\ 
\noalign{\medskip}
15.0 & 5.0 & 4.5 & 0.06 &0.13  \\ 
20.0 & 5.0 & 4.5 & 0.06 &0.12  \\ 
30.0 & 5.0 & 4.5 & 0.05 &0.09  \\ 
\noalign{\medskip}
10.0 & 4.0 & 3.5 & 0.06 &0.13 \\ 
10.0 & 6.0 & 3.5 & 0.08 &0.23 \\ 
\noalign{\medskip}
15.0 & 4.0 & 4.5 & 0.05 &0.09 \\ 

\hline
\end{tabular}
\caption{\label{tab1} The fraction of systems resembling Cygnus X--2 ($f_{\rm cyg}$)
and the fraction of systems which will be located further out
than  Cygnus X--2 ($f_{\rm out}$) as a function
of the initial separation within the post-supernova binary, $d$, the
primary mass {\sl prior} to the supernova, $M_1$, and the mass
of the secondary, $M_2$.}
\end{table}

\section{Summary}

Our binary evolution calculations with full stellar models have verified
that Cygnus X--2 can be understood as the descendant of an
intermediate--mass X--ray binary (IMXB). The most likely evolutionary
solution is an early massive case B sequence, starting from a donor
with mass $\simeq 3.5 \msun$. 
This implies that the presently observed state is rather
short--lived, of order 3 Myr. This in turn points to a large IMXB
formation rate of order $10^{-7} - 10^{-6}$~yr$^{-1}$ 
if Cygnus X--2 is the only such system in the Galaxy. 
It is likely that there are more as yet unrecognized IMXBs, so the
Galactic IMXB formation rate could be even higher.
Hercules X--1 seems to be in the early stages of a similar case A or
case AB evolution (e.g.\ van den Heuvel 1981).
The prehistory of Cygnus X--2 is sensitive to the
width of the main--sequence band in the HR diagram, i.e.\ to
convective overshooting in that phase. The alternative evolutionary
solution for Cygnus X--2 suggested by PR, a case A mass transfer
followed by a case B phase, is viable only if overshooting is very
effective, i.e.\ more effective than hitherto assumed in the
literature.   

Using Mazzitelli's stellar code (Mazzitelli 1989) 
we found that neutron star IMXBs that start case B mass transfer with
initial donor mass $\geq 4 \msun$ will encounter a 
delayed dynamical instability. 
(Note that the maximum donor mass that just avoids this
instability depends on stellar input physics).
The components are likely to merge and
perhaps form a low--mass black hole. Given the high formation rate of
Cygnus X--2--like objects the formation rate of such black holes could
be rather substantial.

We have shown that the large Galactocentric distance of Cygnus X--2
and its high negative radial velocity do not require unusual
circumstances at birth of the neutron star. There are viable
trajectories for Cygnus X--2  that 
firstly cut the Galactic plane inside the solar circle at a time consistent
with the evolutionary age of Cygnus X--2, and 
secondly which require only a moderate neutron star
kick velocity ($<200$km/s). 
We estimate the fraction of systems resembling Cygnus X--2, i.e. of
systems on orbits that reach Galactocentric distances
$>15$~kpc, as $10-20\%$ of the entire IMXB population of systems.

\paragraph*{Acknowledgements} 

MBD gratefully acknowledges the support of a URF from the Royal Society.
ARK thanks the UK Particle Physics \& Astronomy Research Council for a
Senior Fellowship. This work was partially supported by a PPARC
short--term visitors grant. 
We thank the anonymous referee for a careful reading of the manuscript
and for comments that helped to improve the paper.

\bigskip

\appendix{\noindent {\bf APPENDIX: THE GALACTIC POTENTIAL}

\bigskip

The Galactic potential can be modelled as the sum of three
potentials. The spheroid and disc components are given by

\begin{equation}
\Phi_s(R,z) = { G M_s \over \left( R^2 + [a_s + (z^2 + b_s^2)^{1/2}]^2
\right)^{1/2}}
\label{P1}
\end{equation}

\begin{equation}
\Phi_d(R,z) = { G M_d \over \left( R^2 + [a_d + (z^2 + b_d^2)^{1/2}]^2
\right)^{1/2}}
\label{P2}
\end{equation}
where $R^2 = x^2 + y^2$. The component from the Galactic halo
can be derived assuming a halo density distribution, $\rho_{\rm h}$,
given by

\begin{equation}
\rho_h = {\rho_c \over 1 + (r/r_c)^2}
\label{P4}
\end{equation}
where $r^2 = x^2 + y^2 + z^2$. The above density distribution yields
the potential

\begin{equation}
\Phi_h = - {G M_c \over r_c} \left[ {1 \over 2} {\rm ln} \left(
1+ {r^2 \over r_c^2} \right) 
+ {r_c \over r} \ {\rm atan} \left({r \over r_c} \right)
\right]
\end{equation}
where $M_c = 4 \pi \rho_c r_c^3$. The total Galactic potential is
the sum 

\begin{equation}
\Phi=\Phi_s + \Phi_d + \Phi_h	
\end{equation}
Following Paczy\'nski (1990), we use the following choice of parameters:

\begin{equation}
a_s=0, \ b_s = 0.277 \ {\rm kpc}, \  M_s = 1.12 \times 10^{10} M_\odot,  
\end{equation}

\begin{equation}
a_d=3.7 \ {\rm kpc}, \ b_d = 0.20 \ {\rm kpc}, \ 
 M_d = 8.07 \times 10^{10} M_\odot,  
\end{equation}

\begin{equation}
r_c = 6.0 \ {\rm kpc}, \  M_c = 5.0 \times 10^{10} M_\odot, 
\end{equation}
Because of the cylindrical symmetry of the potential, the integration
of the trajectories can be simplified to consider the evolution of
the $z$ and $R$ components only, as given below

\begin{eqnarray}
{dR \over dt} &=& v_R, \ \ {dz \over dt} = v_z, \nonumber \\
{dv_R \over dt} &=& \left( {\partial \Phi \over \partial R} \right)_z
+ { j_z^2 \over R^3}, \ \ {dv_z \over dt} = \left( {\partial \Phi 
\over \partial z } \right)_R
\label{P9}
\end{eqnarray}
where the $z$ component of the angular momentum, $j_z = R v_\phi$.



\end{document}